\documentstyle[12pt]{article} 
\begin{document}
\baselineskip 16pt plus 2pt minus 2pt
\newcommand{\beq}{\begin{equation}}
\newcommand{\eeq}{\end{equation}}
\newcommand{\beqa}{\begin{eqnarray}}
\newcommand{\eeqa}{\end{eqnarray}}
\newcommand{\dfrac}{\displaystyle \frac}
\renewcommand{\thefootnote}{\#\arabic{footnote}}
\newcommand{\ve}{\varepsilon}
\newcommand{\krig}[1]{\stackrel{\circ}{#1}}
\newcommand{\barr}[1]{\not\mathrel #1}

\begin{titlepage}
 
\noindent REVISED AND ENLAREGD VERSION \hfill KFA-IKP(Th)-1996-14

\hfill LPT--96--22

\hfill hep-ph/9611253


\vspace{1.0cm}

\begin{center}

{\large  \bf {
Aspects of chiral pion--nucleon physics}}

\vspace{1.2cm}
                              
{\large V. Bernard$^{\ddag}$,
N. Kaiser$^{\diamond}$, 
Ulf-G. Mei\ss ner$^{\dag}$}

\vspace{1.0cm}

$^{\ddag}$Universit\'e Louis Pasteur, Laboratoire de Physique
Th\'eorique\\ BP 28, F--67037 Strasbourg, France\\
{\it email: bernard@crnhp4.in2p3.fr} \\

\vspace{0.4cm}
$^{\diamond}$Technische Universit\"at M\"unchen, Physik Department T39\\ 
James-Franck-Stra{\ss}e, D--85747 Garching, Germany\\
{\it email: nkaiser@physik.tu-muenchen.de}\\

\vspace{0.4cm}
$^{\dag}$FZ J\"ulich, IKP (Theorie), D--52425 J\"ulich, Germany\\
{\it email: Ulf-G.Meissner@kfa-juelich.de} \\

\end{center}

\vspace{0.4cm}

\begin{abstract}
\noindent The next--to--leading order chiral pion--nucleon Lagrangian 
contains seven finite low--energy constants. Two can be fixed from the
nucleon anomalous magnetic moments and another one from the quark mass
contribution to the neutron--proton mass splitting. We find a set of
nine observables, which to one loop order do only depend on the
remaining four dimension two couplings. These are then determined from
a best fit. We also show that their values can be understood in terms
of resonance exchange related to $\Delta$ excitation as well as
vector and scalar meson exchange. In particular, we discuss the role of the
fictitious scalar--isoscalar meson. We also investigate the chiral expansion of
the two P--wave scattering volumes $P_1^-$ and $P_2^+$ as well as the isovector
S--wave effective range parameter $b^-$. The one--loop calculation is in good 
agreement with the data. The difference  $P_1^- - P_2^+$ signals chiral loop 
effects in the $\pi N$ P--waves. The calculated D-- and F--wave 
threshold parameters compare well with the empirical values. 

\end{abstract}


\vspace{2cm}

\centerline{Accepted for publication in Nuclear Physics A}

\vfill

\end{titlepage}


\section{Introduction and summary}
\label{sec:intro}

Chiral perturbation theory is the tool to systematically investigate the
consequences of the spontaneous and explicit chiral symmetry breaking in 
QCD. S--matrix elements and transition currents of quark operators are
calculated with the help of an effective field theory formulated in terms
of asymptotically observed fields, the Goldstone bosons and the low--lying
baryons. A systematic perturbative expansion in terms of small external
momenta and meson masses is possible. We call this double expansion from here
on chiral expansion and denote the small parameters collectively by
$q$. Beyond leading order, coupling constants not fixed by chiral symmetry
appear, the so--called low--energy constants (LECs). For the chiral pion
Lagrangian, i.e. the two--flavor case, these were determined more than a 
decade ago by Gasser and Leutwyler \cite{gl84} by fitting a set of 
observables calculated at next--to--leading order. In the presence of 
nucleons, the situation is less satisfactory. At next--to--leading order
($q^2$), seven finite LECs appear \cite{bkkm} and 24  at order $q^3$ 
\cite{eckerm}, which is the first order where loops can contribute
(11 of these are finite, the other 13 are scale--dependent because they are
needed to absorb the one--loop divergences). The dimension two pion--nucleon
Lagrangian can be written as
\beq \label{leff2}
{\cal L}_{\pi N}^{(2)} = \sum_{i=1}^7 c_i \, O_i \,\,\, ,
\eeq
with the $O_i$ monomials in the fields of dimension two. At present, 
no completely systematic evaluation of the LECs $c_i$ exists. In particular,
the four LECs called $c_{1,2,3,4}$ related to pion--nucleon scattering
have been determined to one loop accuracy in the review \cite{bkmrev}
and to order ${\cal O}(q^2)$ in \cite{bkmpipin}. The resulting values
differ by factors of 1.5. None of these determinations is satisfactory
since some of the input data are not very well known or large cancelations
between individual terms appear (the best example is the isoscalar $\pi N$
S--wave scattering length $a^+$). Furthermore, if one wants to extract the
dimension three LECs, one needs the $c_i$ as input since they enter  via
$1/m$ suppressed vertices at that order (compare the form of the complete
${\tilde {\cal L}}_{\pi N}^{(3)}$ in \cite{eckerm}). Clearly, a more 
stringent determination of these parameters is called for. A reliable 
determination should also be based on more observables than LECs in order to 
have some consistency checks. We close this gap
in this paper. Without going into details, we will proceed as follows.
The LECs $c_6$ and $c_7$ can be directly inferred from the anomalous magnetic
moments of the proton and the neutron \cite{bkkm}. In the absence of
external (pseudo)scalar fields, the operator $O_5$ is
only non--vanishing for unequal light quark masses, $m_u \ne m_d$. The
corresponding LEC $c_5$ can be extracted from the strong contribution to the
neutron--proton mass difference. For the other four coupling constants,
we find a set of nine observables which at one--loop order are
given entirely in terms of tree graphs with insertions from 
${\cal L}_{\pi N}^{(1)} + {\cal L}_{\pi N}^{(2)}$ together with their
$1/m$-corrections and finite loop contributions. These are very special cases
since in general at this order divergences would appear and thus dimension
three LECs would be needed. From a best fit to these nine observables, we are
able to determine the LECs $c_{1,2,3,4}$. 

\medskip

Furthermore, in the meson sector it can be shown
that the numerical values of the renormalized LECs $L_i^r (\mu = M_\rho)$
can be understood to a high degree of accuracy from resonance saturation,
i.e. they can be expressed in terms of resonance masses and coupling 
constants of the low--lying vector ($V$), axial--vector ($A$), scalar ($S$)
and pseudoscalar ($P$) multiplets (the $\eta '$, to be precise) \cite{reso} 
(in some cases, there is
also some contribution from tensor mesons \cite{dt}). We  investigate
how well one can understand the numerical values of the $c_i$ in terms
of baryonic ($\Delta, N^*, \ldots$) and mesonic ($S,V,\ldots$) excitations.
In particular, we discuss the role of the fictitious scalar--isoscalar meson
and show how such correlated two--pion exchange reveals itself in certain
LECs. Since we do not include the $\Delta$ as an active degree of freedom 
in the effective field theory, it contributes dominantly to some of the
LECs as it is expected from  the important role this resonance plays in
pion--nuclear physics \cite{ew}. We have already shown in a series of
detailed calculations concerning a variety of reactions in the
corresponding threshold regions that it is
legitimate to encode the effects of the $\Delta$ in the pertinent LECs,
see the review \cite{bkmrev}. Since here we mostly consider threshold
parameters (like scattering lengths and effective ranges), this procedure
is expected to be sufficiently accurate. It remains to be proven by
the authors who include the $\Delta$ as an active degree of freedom that
their approach is equally precise (in the threshold region, of course).

\medskip

The pertinent results of this investigation can be summarized as follows:

\begin{enumerate}

\item[(i)] We have determined the seven finite low--energy constants
of the dimension two chiral pion--nucleon Lagrangian, ${\cal L}_{\pi
N}^{(2)}$. We have found a set of nine observables that to one--loop
order $q^3$ are given entirely in terms of tree graphs including
insertions $\sim c_1, c_2, c_3, c_4$ and finite loop contributions,
but with none from the 24 new LECs of ${\cal L}_{\pi N}^{(3)}$. A best fit
allows to pin down these LECs. The other three can be determined from the
strong neutron--proton mass difference ($c_5$, which is only relevant in the
case $m_u \neq m_d$) and from the anomalous magnetic moments of the
proton and the neutron ($c_6$, $c_7$). The resulting values are listed
in table~1 in section~4.

\item[(ii)] We have shown that the empirical values of the LECs $c_1 ,
\ldots , c_4$ can be understood from resonance exchange. Assuming that
$c_1$ is saturated completely by scalar meson exchange, the values for
$c_2$, $c_3$ and $c_4$ can be understood  from a
combination of $\Delta$, $\rho$ and scalar meson exchange.
It is remarkable that the scalar mass to coupling constant
ratio $M_S / \sqrt{g_S}$ needed to saturate the LEC $c_1$ is in
perfect agreement with typical ratios obtained in boson--exchange
models of the NN force, where the $\sigma$--meson models the strong
pionic correlations coupled to nucleons. 
There is, however, some sizebale uncertainty related to the $\Delta$
contribution as indicated by the ranges given in table~1. 
Concerning the LECs
$c_6$ and $c_7$ related to $\krig{\kappa}_v$ and $\krig{\kappa}_s$, we
find that the isoscalar and isovector anomalous moments in the chiral limit 
can be well understood from neutral vector meson exchange. For the LEC $c_5$, 
resonance saturation can not be used
since there is no information on isospin--violating coupling constants. 

\item[(iii)] Having established that resonance saturation can explain
the LECs related to pion--nucleon scattering, we have considered the
chiral expansion of the P--wave scattering volumes $P_1^-$ and $P_2^+$
to order $q^3$. After renormalizing the appearing divergences, the
chiral predictions agree at the few percent level with the empirical values. 
The largest uncertainty comes actually from the $\Delta(1232)$-contribution. 
The difference $P_1^- - P_2^+$ shows the relevance of chiral
loops in the $\pi N$ P--waves. 

\item[(iv)] The eight D-- and F--wave threshold parameters $a_{l\pm}^\pm$
($l = 2, 3$) are given to order $q^3$ by lowest order tree and loop
graphs only. The calculated values agree nicely with the empirical ones.

\end{enumerate}

\medskip

This investigation is the  first systematic attempt to pin down the
low--energy constants of the chiral pion--nucleon Lagrangian. Clearly,
more precise data are needed to sharpen the determination of the
$c_i$. The present work, however, paves the way of fixing a subset of
the dimension three LECs enumerated in \cite{eckerm}. For that, a
systematic study of $\pi N$ scattering to order $q^3$ should be performed.
Such a study has recently been performed by Moj\v zi\v s \cite{mm}.


\section{Effective Lagrangian at next--to--leading order}

In this section, we briefly review the next--to--leading order 
pion--nucleon Lagrangian ${\cal L}_{\pi N}^{(2)}$
to fix our notation. We work in the path
integral formulation of heavy baryon chiral perturbation theory which
automatically obeys reparametrization invariance. All details are
spelled out in \cite{bkkm} or the review \cite{bkmrev}. The pions are
collected in the SU(2) matrix $U(x) = u^2 (x)$ and the proton and the
neutron in the iso--doublet $N(x)$. With $v_\mu$
the four--velocity of the heavy nucleon fields and $S_\mu$ the covariant
spin--operator \`a la Pauli--Lubanski, ${\cal L}_{\pi N}^{(2)}$ takes the
form
\begin{eqnarray}  \label{LpiN2}
{\cal L}_{\pi N}^{(2)} =  && {\bar {N}}\biggl\{ \frac{1}{2\krig{m}}  (v\cdot
D)^2-\frac{1}{2\krig{m}}{D \cdot D} 
-\frac{i\, \krig{g}_A}{2\krig{m}}\{ S\cdot D , v \cdot u \} \nonumber \\
& &\, + c_1 \,{\rm Tr}(\chi_+) 
+ \biggl(c_2-\frac{\krig{g}_A^2}{8\krig{m}}\biggr)(v \cdot u)^2
+ c_3 \, u \cdot u \nonumber \\
& &+ \biggl(c_4+\frac{1}{4\krig{m}}\biggr)[S^\mu,S^\nu]u_\mu u_\nu
+ c_5 \biggl( \chi_+ - \frac{1}{2}{\rm Tr}(\chi_+) \biggr) \nonumber \\
& &-\frac{i}{4\krig{m}}[S^\mu , S^\nu ]\, 
\biggl[(1+\krig{\kappa}_v) \, f_{\mu\nu}^+ 
+ 2(1+ \krig{\kappa}_s)\,v_{\mu\nu}^{(s)} \biggr] 
\biggr\} {N} \,\, ,
\end{eqnarray}
with
\begin{eqnarray}
\chi_+ &=& u^\dagger \chi u^\dagger + u \chi^\dagger u \,\,\, , \nonumber \\
f_{\mu \nu}^+ &=&  u F_{\mu \nu}^L  u^\dagger + u^\dagger F_{\mu \nu}^R  u
\,\,\, , \nonumber \\ u_\mu &=& i \, u^\dagger \nabla_\mu U u^\dagger \,\,\, .
\end{eqnarray}
Here, $F_{\mu \nu}^{L,R}$ are the non--abelian field strength tensors
of external left/right handed vector gauge fields  and
$v^{(s)}_{\mu \nu}$ is defined analogously in terms of the isosinglet vector
field $v_\mu^{(s)}$ necessary to generate the full electromagnetic current.  
$D_\mu$ is the covariant derivative acting on the nucleons and,
similarly, $\nabla_\mu$ the one acting on the pions. Furthermore, $\chi = 2 B_0
{\cal M}+\dots$ with ${\cal M}$=diag($m_u,m_d$) the light quark mass matrix
and $B_0 = |\langle 0 |\bar{u}u |0\rangle|/F_\pi^2$, with $F_\pi
=92.4\,$ MeV the pion decay constant. Some of the terms in eq.(\ref{LpiN2})
receive $1/m$ corrections from the expansion of the relativistic Dirac
$\pi N$ Lagrangian. We have kept these explicitly one reason being that a
phenomenological interpretation in terms of resonance exchange can not
generate such terms. All parameters appearing are taken to be at their
values in the chiral limit, i.e.
\beq
\krig Q = Q \, [ 1 + {\cal O}(m_q^\alpha) ] \,\,\, ,
\eeq
where $m_q$ denotes any one of the light quark masses or its average. 
In most cases, one has $\alpha=1/2$, exceptions being the anomalous isoscalar
magnetic moment $\krig{\kappa}_s$ 
and $c_5$ with $\alpha=1$ (see below). In what follows, we can
identify the nucleon mass and the axial--vector coupling constant
with their physical values, 
$\krig{m} = m_p=938.27\,$MeV and $\krig{g}_A = g_A =1.26$.
We will now be concerned with the numerical
values of the LECs appearing in  ${\cal L}_{\pi N}^{(2)}$, these are
the $c_i$ $(i=1, \ldots ,5)$ as well as $\krig{\kappa}_s$ and 
$\krig{\kappa}_v$. The machinery to do these calculations is spelled
out in detail in \cite{bkmrev}.


\section{Calculation of observables}

In this section, we calculate various observables to pin down the LECs
$c_i$. The $c_{1,2,3,4}$ are all related to pion--nucleon threshold and
subthreshold parameters and the much discussed pion--nucleon $\sigma$--term.
We consider here only observables which to one loop order ${\cal O}(q^3)$
are given by tree graphs including the $c_i$ and finite loop corrections
but have no contribution from the 24 LECs of ${\cal L}_{\pi N}^{(3)}$. 

Consider first a subset of observables which depend on the LECs $c_1$,
$c_2$ and $c_3$. We introduce the small parameter $ \mu = M_\pi /m $,
i.e. the pion to nucleon mass ratio. Our notation concerning the
$\pi N$ amplitudes and parameters is identical to the one used by
H\"ohler \cite{hoeh}. 
Calculation of the $\sigma$--term and the isospin--even
scattering amplitude at and below threshold gives four relations (the
one--loop contributions to the $\pi N$ scattering amplitude are collected
in app.~A),
\begin{eqnarray} 
\label{fourr}
\sigma(0)  &=&  -4 c_1 M_\pi^2 -{9g_{\pi N}^2 M_\pi^3 \over 64
\pi m^2 }  + {\cal O}(M_\pi^4)  \,\,\, ,  \\
a_{00}^+ &=& {2M_\pi^2 \over  F_\pi^2 } \Big(c_3-2c_1 \Big) +
{g_A^2 M_\pi^3  \over 8 \pi F_\pi^4 } \Big(g_A^2+{3\over 8} \Big) + {\cal
O}(M_\pi^4)  \,\,\, ,  \\ 
a_{01}^+ &=& -{c_3\over  F_\pi^2 } - {g_A^2  M_\pi
\over 16 \pi F_\pi^4 } \Big(g_A^2+{77\over 48} \Big) + {\cal
O}(M_\pi^2) \,\,\, , \\ 
d_{10}^+ &=& {2c_2\over F_\pi^2 } - {M_\pi \over 8 \pi F_\pi^4 }
\Big({5\over 4} g_A^4+1 \Big) + {\cal O}(M_\pi^2)  \, \,\, . 
\end{eqnarray}
Note that the formula for $a_{01}^+$ was already derived in \cite{bkmrev} for
the so--called axial polarizability $\alpha_A = 2 a_{01}^+$. 
Another relation can be derived from the isospin--even non--spin--flip
scattering volume, $P_1^+$,
\begin{eqnarray}
\label{P1+}
P_1^+ &=& {4 \pi \over 3}  (1 + \mu) \Big( 4a_{33}+2a_{31}+
2a_{13}+a_{11} \Big)\nonumber \\ &=& {2\over F_\pi^2 }\Big(c_2\,
\mu -c_3 \Big) +{g_{\pi N}^2 \mu \over 4 m^3 }  - {g_A^2
M_\pi \over 12 \pi F_\pi^4 } \Big(g_A^2+{77\over 32} \Big) + {\cal O}(M_\pi^2)
\, \, \,  .
\end{eqnarray}
Consider also the real part of isospin--even $\pi N$ forward amplitude
close to threshold,
\beqa \label{expTp}
{\rm Re} \, T^+(\omega) &=& 4\pi \frac{\sqrt{s}}{m} \, (a^+ + b^+ {\vec q\,}^2)
+ P_1^+ \,  {\vec q\,}^2 + {\cal O}({\vec q\,}^4) \nonumber \\
&=& T^+ (M_\pi) + {\vec q\,}^2 \, \beta^+ + {\cal O}({\vec q\,}^4) 
\,\, ,
\eeqa
with 
\begin{equation} \label{betap} 
\beta^+ = \frac{1}{2 M_\pi} \frac{\partial}{\partial \omega}
{\rm Re} \, T^+ (\omega) \bigl|_{\omega=M_\pi} =
P_1^+ + 4\pi (1+\mu) \Big( b^+ + {a^+\over 2 m
M_\pi} \Big) \,\, .
 \end{equation}
Here, $\omega$ denotes the pion cms energy.
The chiral expansions of the scattering length $a^+$ \cite{bkmpin} 
and of the range parameter $\beta^+$ take the form
\begin{equation} 
T^+(M_\pi)  = 4 \pi(1+\mu) a^+ ={M_\pi^2 \over F_\pi^2 }\Big (
-4c_1+ 2c_2 -{g_A^2 \over 4m} +2c_3 \Big) +{3g_A^2 M_\pi^3 \over 64 \pi F_\pi^4
} + {\cal O}(M_\pi^4) \, ,
\end{equation}  
\begin{equation}
 \beta^+ = {2c_2\over F_\pi^2 } (1 + 2 \mu )
+ {g_{\pi N}^2 \over 4 m^3} ( 1
+ 2\mu) - {g_A^4 M_\pi \over 12 \pi F_\pi^4 } + {\cal O}(M_\pi^2)  \, .
\end{equation}
The calculation leading to these results is somewhat tricky. The tree
terms are most easily evaluated by considering the relativistic 
pion--nucleon Lagrangian with the two couplings $c_2 '$ and $c_2 ''$,
see ref.\cite{bkmpipin}. It leads to the forward scattering amplitude
\begin{equation} 
T^+  = (c_2 ' + c_2 '') \, {s -m^2 -M_\pi^2 \over 2 m^2 F_\pi^2 } =
2c_2 \, {\omega_L^2 \over F_\pi^2} \,\,\, ,
\end{equation}  
with $\omega_L$ the pion laboratory energy. Expanding in powers of
$1/m$ gives the desired result.

\medskip

The LEC $c_4$ appears in the chiral expansion of the isospin--odd spin--flip
scattering volume $P_2^-$ and the subthreshold parameter $b_{00}^-$,
\begin{eqnarray}
P_2^- &=& {4 \pi\over 3}(1+ \mu) \Big(
a_{33}-a_{31}-a_{13}+a_{11} \Big) \nonumber \\ 
&=& {2+\mu \over 8 m F_\pi^2}  + {c_4 \over F_\pi^2} \Big(1 +
\mu\Big) - {g_A^2 M_\pi \over 48 \pi F_\pi^4 } \Big( 2g_A^2 +3 \Big) + {\cal
O}(M_\pi^2)  \, , \\
b_{00}^- &=& {1\over 2 F_\pi^2 } \Big(1+4m c_4 \Big) - {g_A^2 m
M_\pi \over 8 \pi F_\pi^4 } \Big(1+g_A^2 \Big) + {\cal O}(M_\pi^2)\,\, . 
\end{eqnarray} 

\medskip

In the absence of a precise scheme to separate isospin--violating
quark mass effects from the ones of virtual photons for dynamical
processes, we use the information on the strong contribution to the
neutron--proton mass difference to pin down $c_5$,
\begin{equation} \label{mnpstr}
(m_n-m_p)^{(\rm non-elm)} = 4 c_5 \, B_0 \, (m_u - m_d) + {\cal O}(M_\pi^4) =
4 c_5 M_\pi^2 {m_u - m_d \over m_u + m_d} + {\cal O}(M_\pi^4)  \, \, .  
\end{equation}
We assume here the standard scenario of spontaneous chiral symmetry
breaking, i.e. $B_0 \gg F_\pi$. We remark that the generalized
scenario with $B_0 \sim F_\pi$ would lead to a vastly different value
of $c_5$.  Other observables sensitive to this LEC are the $\pi N$
S-wave scattering lengths (taken not in the isospin limit) for processes
involving at least one neutral pion \cite{weinmit}.

\medskip

The anomalous magnetic moments appearing in the dimension two Lagrangian
have been calculated in \cite{bkkm,bkmrev},
\begin{eqnarray} 
\krig{\kappa}_s &=&  \kappa_s + {\cal O}(M_\pi^2) \,\, , \\
\krig{\kappa}_v &=&  
\kappa_v + {g_{\pi N}^2 \mu \over 4 \pi} + {\cal O}(M_\pi^2) \, \, .
\end{eqnarray}
These are related to the LECs $c_6$ and $c_7$ used there via
\beq
c_6 = \krig{\kappa}_v \,\, , \quad 
c_7 = \frac{1}{2}(\krig{\kappa}_s - \krig{\kappa}_v ) \,\, .
\eeq
There are no one--loop corrections at order $q^3$ to $\kappa_s$ since
the spectral functions of the isoscalar electromagnetic form factors 
start at the three--pion cut, $t_0 = 9 M_\pi^2$.


\section{Determination of the low--energy constants}

First, we must fix parameters. We use $ g_{\pi N} = 13.4$ and
$g_A$ as determined from the Goldberger--Treiman relation,
$g_A = g_{\pi N} F_\pi /m = 1.32$. We also have performed fits with
the smaller $ g_{\pi N} = 13.05$ and thus $g_A=1.29$. For the
$\sigma$--term, we use $ \sigma(0)= 45 \pm 8 \,$MeV \cite{gls}. The
threshold and subthreshold parameters we take from \cite{hoeh}, these
are $a_{00}^+ = (-1.46 \pm 0.10) \, M_{\pi}^{-1}$, 
$b_{00}^- =  (10.36 \pm 0.10) \, M_{\pi}^{-2}$, 
$a_{01}^+ =  (1.14 \pm 0.02) \, M_{\pi}^{-3}$, 
$d_{10}^+ =  (1.12 \pm 0.02) \, M_{\pi}^{-3}$, 
$P_1^+ = (3.01 \pm 0.05) \, M_{\pi}^{-3}$, 
$P_2^-  = (1.00 \pm 0.02 ) \, M_\pi^{-3}$ and
$b^+ = -(44 \pm 7) \cdot 10^{-3}\, M_\pi^{-3}$. 
For the isoscalar S--wave scattering length, we use a generous bound 
$a^+  =( 0 \pm 10 ) \cdot 10^{-3} \, M_{\pi}^{-1}$ since
the Karlsruhe--Helsinki phase shifts \cite{koch} give 
$ a^+ = -( 8.3 \pm 3.8 ) \cdot 10^{-3} \, M_{\pi}^{-1}$ where as the
new PSI-ETHZ \cite{leisi} value is small and positive,   
$ a^+  =(0 \ldots 4 ) \cdot 10^{-3} \, M_{\pi}^{-1}$. Consequently, the
value for $\beta^+$ follows to be $\beta^+ = (2.36 \pm 0.15) \,
M_{\pi}^{-3}$, adding the uncertainties of $P_1^+,\,b^+$ and $a^+$ in
quadrature. The magnetic moments are known very precisely, for our
purpose it suffices to take $\kappa_v = 3.706$ and $\kappa_s = -0.120$.
Finally, we need a value for the strong neutron--proton mass
difference. This has been evaluated in great detail in \cite{glpr} and
we thus use $ (m_n-m_p)^{(\rm non-elm)}  = (2.0 \pm 0.3) \, {\rm
MeV}$. We remind the reader that the photon cloud contribution as
calculated via the Cottingham formula is about 0.8 MeV. The light quark mass
ratio has been determined recently by Leutwyler \cite{leutw}, $m_d/m_u =
1.8$.  

\medskip

With  the error bars for the various observables as given above, we
obtain as values of the $c_i$ 
\beq
c_1 = -1.02 \pm 0.06 \, {\rm GeV}^{-1} \,\, , \,\,
c_2 =  3.32 \pm 0.03 \, {\rm GeV}^{-1} \,\, , \,\,
c_3 = -5.57 \pm 0.05 \, {\rm GeV}^{-1} \,\, .
\eeq 
for our central set of
parameters. The uncertainties for $c_{1,2,3}$ refer to the parabolic
errors of the MINUIT fitting routine used. We remark that the
fit prefers a negative value for $a^+$ and the $\sigma$--term on the 
large side, $a^+ = -10.1 \cdot 10^{-3} \, M_\pi^{-1},\, \sigma(0) = 54.9$ MeV. 
Clearly, the $\chi^2$/dof of 3.03 shows that the input
data are not all mutually consistent (to order $q^3$). Higher order
corrections not yet calculated might remove these discrepancies. 
In particular, almost half of
the total $\chi^2$ stems from $P_1^+$, i.e. the error in $c_2$ and $c_3$ is
certainly larger than the one obtained from the fitting procedure. 
To get a more realistic estimate of the uncertainties for the various LECs, 
we have performed a fit were we have increased the uncertainties in all
observables to $\pm 15\%$ leaving $\sigma(0)$ and the range for $a^+$
as before. Considering the present status of the low--energy
pion--nucleon scattering data basis, we consider such uncertainties
as more realistic. For that fit, the $\chi^2$/dof = 0.33 is much better and
the resulting values are $c_1 = -0.93 \pm 0.09 \, {\rm GeV}^{-1}$,  
$c_2 =  3.34 \pm 0.18 \, {\rm GeV}^{-1}$, and $c_3 = -5.29 \pm 0.25 
\, {\rm GeV}^{-1}$.
These we consider our central values as given in table~1 
(the uncertainties are rounded towards the larger side) together with
the dimensionless couplings $c_i ' = 2 m c_i$, $i=1,\ldots,5$ (the prefactor $2m$ 
appears naturally in the heavy mass expansion). This fit leads to
$\sigma (0) = 47.6\,$MeV and $a^+ = -4.7 \cdot 10^{-3} \,
M_\pi^{-1}$. For comparison, the values determined in the review \cite{bkmrev}
based solely on the input from the $\sigma$-term, $a_{01}^+$ and $a^+$
from the Karlsruhe--Helsinki analysis, are 
$c_1 = -0.87 \pm 0.11 \, {\rm GeV}^{-1}$,  
$c_2 =  3.34 \pm 0.27 \, {\rm GeV}^{-1}$ and 
$c_3 = -5.25 \pm 0.22 \, {\rm GeV}^{-1}$.
 If we use the smaller value for $g_{\pi N} = 13.05$ (i.e. $g_A
= 1.29$ from the GTR), we get $c_1 = (-1.01 \pm 0.06) \, {\rm
GeV}^{-1}$, $c_2 = (3.20 \pm 0.03) \, {\rm GeV}^{-1}$ and
$ c_3 = (-5.45 \pm 0.05) \, {\rm GeV}^{-1}$. The $\chi^2$/dof = 3.74 is 
considerably worse. This is, however, not due to one observable but
almost all of them contribute more to the total $\chi^2$ compared to
the choice $g_{\pi N} = 13.4$. Again, for the enlarged uncertainties
one gets a substantially lower $\chi^2$/dof = 0.30 for the values
$c_1 = -0.91 \pm 0.09 \, {\rm GeV}^{-1}$, $c_2 =  3.25 \pm 0.18 \, 
{\rm GeV}^{-1}$, and $c_3 = -5.16 \pm 0.25 \, {\rm GeV}^{-1}$. 
Note that the tree level
prediction $b_{00}^- = 2m\, P_2^-$ is violated by $30\%$. With the 
inclusion of loop effects,  however, a consistent value of $c_4$ can 
be obtained from both observables. The same is true for the set of seven
observables depending on $c_{1,2,3}$. An omission of the loop corrections 
results in a ten times larger $\chi^2$/dof.  For the observables considered
here, the loop effects are typically of the order of 30\% to 50\%, i.e. not
small. It is also worth emphazising that we do not quote an
uncertainty for $c_6$ and $c_7$ since the magnetic moments of the
proton and the neutron have been determined with extreme precision.
Notice that in \cite{weinmit} a somewhat larger value for $c_5$ is obtained
based on an leading order SU(3) estimate for $(m_n-m_p)^{\rm non-elm}$.

\renewcommand{\arraystretch}{1.3}

\begin{center}

\begin{tabular}{|l|r|r|r|c|}
    \hline
    $i$         & $c_i \quad \quad$   &  $c_i ' \quad \quad$  &
                  $ c_i^{\rm Res} \,\,$ cv & 
                  $ c_i^{\rm Res} \,\,$ ranges    \\
    \hline
    1  &  $-0.93 \pm 0.10$  & $-1.74 \pm 0.19$ & $-0.9^*$ & -- \\
    2  &  $3.34  \pm 0.20$  & $6.27  \pm 0.38$ & $3.9\,\,$ & $2 \ldots 4$ \\    
    3  &  $-5.29 \pm 0.25$  & $-9.92 \pm 0.47$& $-5.3\,\,$ 
                                     & $-4.5 \ldots -5.3$ \\
    4  &  $3.63  \pm 0.10$   & $6.81  \pm 0.19$ & $3.7\,\,$ 
                                     & $3.1 \ldots 3.7$ \\    
    5  &  $-0.09 \pm 0.01$   & $-0.17 \pm 0.02$ & $-\,\,$ & $-\,\,$ \\    
    \hline
    6           &  $5.83$             & --  & $6.1\,\,$  & $-\,\,$ \\
    7           &  $-2.98$            & --  & $-3.0\,\,$ & $-\,\,$ \\
    \hline
  \end{tabular}

\smallskip 

\end{center}

\noindent Table 1: Values of the LECs $c_i$ in GeV$^{-1}$
and the dimensionless couplings $c_i '$
for $i=1,\ldots,5$. The LECs $c_{6,7}$ are dimensionless.
Also given are the central values (cv) and the ranges
for the $c_i$ from resonance
exchange as detailed in section~5. The $^*$ denotes an input quantity.


\bigskip


\section{Phenomenological interpretation of the low-energy constants}

In this section, we will be concerned with the phenomenological
interpretation of the values for the LECs $c_i$. For that, guided by
experience from the meson sector \cite{reso}, we use resonance
exchange. To be specific, consider an effective Lagrangian with
resonances chirally coupled to the nucleons and pions. One can
generate local pion--nucleon operators of higher dimension with given
LECs by letting the resonance masses become
very large with fixed ratios of coupling constants to masses. That
procedure amounts to decoupling the resonance degrees of freedom from
the effective field theory. However, the traces of these frozen
particles are encoded in the numerical values of certain LECs. In the
case at hand, we can have baryonic ($N^*$) and mesonic ($M$)
excitations,
\beq \label{cirdef}
c_i = \sum_{N^*=\Delta,R,\ldots} c_i^{N^*} + \sum_{M=S,V,\ldots} c_i^M
\,\,\, ,
\eeq
where $R$ denotes the Roper $N^* (1440)$ resonance. We remark again that the $c_i$
are finite and scale--independent.

\medskip

We consider first scalar ($S$) meson exchange. The SU(2) $S\pi\pi$
interaction can be written as
\beq
{\cal L}_{\pi S} = S \, \Big[ \bar{c}_m \, {\rm Tr}(\chi_+) + 
\bar{c}_d \, {\rm Tr} (u_\mu u^\mu) \Big] \,\,\, .
\eeq
{}From that, one easily calculates the s--channel scalar meson
contribution to the invariant amplitude $A(s,t,u)$ for elastic 
$\pi \pi$ scattering,
\beq
A^S (s,t,u) = \frac{4}{F_\pi^4 (M_S^2 -s)} \, [ 2 \bar{c}_m 
M_\pi^2 + \bar{c}_d (s- 2M_\pi^2) ]^2 +{16 \bar c_m M_\pi^2 \over 3
F_\pi^4 M_S^2 } \Big[ \bar c_m M_\pi^2 + \bar c_d(3s-4M_\pi^2) \big]   \,\,\, .
\eeq
Comparing with the SU(3) amplitude calculated in \cite{bkmsu3}, we
are able to relate the $\bar{c}_{m,d}$ to the  ${c}_{m,d}$ of
\cite{reso} (setting $M_{S_1} = M_{S_8} =M_S$ and using the large--$N_c$
relations $\tilde{c}_{m,d} = c_{m,d} / \sqrt{3}$ to express the
singlet couplings in terms of the octet ones),
\beq
\bar{c}_{m,d} = \frac{1}{\sqrt{2}} \, c_{m,d} \quad ,
\eeq
with $|c_m| =42\,$MeV and $|c_d| = 32\,$MeV \cite{reso}. Assuming now
that $c_1$ is entirely due to scalar exchange, we get
\beq 
c_1^S = - \frac{g_S \, \bar{c}_{m}}{M_S^2} \quad .
\eeq
Here, $g_S$ is the coupling constant of the scalar--isoscalar meson to
the nucleons, ${\cal L}_{SN} = - g_S \, \bar{N} N \, S$. What this 
scalar--isoscalar meson is essentially doing is to mock up the strong
pionic correlations coupled to nucleons. Such a phenomenon 
is also observed in the meson sector. The one loop description
of the scalar pion form factor fails beyond energies of 400 MeV,
well below the typical scale of chiral symmetry breaking,
$\Lambda_\chi \simeq 1\,$GeV. Higher loop effects are needed
to bring the chiral expansion in agreement with the data \cite{game}.
Effectively, one can simulate these higher loop effects by introducing
a scalar meson with a mass of about 600 MeV. This is exactly the line
of reasoning underlying the arguments used here (for a pedagogical 
discussion on this topic, see \cite{cnpp}). It does, however, not mean
that the range of applicability of the effective field theory is
bounded by this mass in general. In certain channels with strong pionic
correlations one simply has to work harder than in the channels
where the pions interact weakly (as demonstrated in great detail
in \cite{game}) and go beyond the one loop approximation
which works well in most cases. For $c_1$ to be 
completely saturated by scalar exchange, $c_1 \equiv c_1^S$, we need
\beq
\frac{M_S}{\sqrt{g_S}} = 180 \, {\rm MeV} \quad .
\eeq
Here we made the assumption that such a scalar has the same couplings
to pseudoscalars as the real $a_0 (980)$ resonance.
It is interesting to note that the effective $\sigma$--meson in the 
Bonn one--boson--exchange potential \cite{bonn} with $M_S = 550\,$MeV and
$g_S^2/(4 \pi) = 7.1$ has $M_S / \sqrt{g_S} = 179\,$MeV. This number
is in stunning agreement with the the value demanded from scalar meson
saturation of the LEC $c_1$. With that, the scalar meson contribution
to $c_3$ is fixed including the sign, since $c_m c_d >0$ (see ref.\cite{reso}),
\beq
c_3^S = -2 \frac{g_s \, \bar{c}_d}{M_S^2} = 2 \frac{c_d}{c_m} \, c_1 
= -1.40 \, {\rm GeV}^{-1} \quad .
\eeq
The isovector $\rho$ meson only contributes to $c_4$. Taking a universal
$\rho$--hadron coupling and using the KSFR relation, we find
\beq
c_4^\rho = \frac{\kappa_\rho}{4m} = 1.63 \, {\rm GeV}^{-1} \quad ,
\eeq
using $\kappa_\rho = 6.1 \pm 0.4$ from the analysis of the nucleon
electromagnetic form factors, the process $\bar{N}N \to \pi \pi$
\cite{mmd} \cite{hoehpi} and the phenomenological one--boson--exchange
potential for the NN interaction.

\medskip

We now turn to the baryon excitations. Here, the dominant one is the 
$\Delta (1232)$. Using the isobar model and the SU(4) coupling
constant relation (the dependence on the off--shell parameter $Z$ has 
already been discussed in \cite{bkmrev}), the $\Delta$ contribution to
the various LECs is readily evaluated,
\beq \label{delta}
c_2^\Delta = -c_3^\Delta = 2 c_4^\Delta = \frac{g_A^2 \, (m_\Delta
-m)}{2 [(m_\Delta -m)^2 - M_\pi^2]} = 3.83 \, {\rm GeV}^{-1} \,\, .
\eeq
These numbers we consider as our central values.
Unfortunately, there is some sizeable uncertainty in these $\Delta$
contributions. Dropping e.g. the factor $M_\pi^2$ in the denominator
of eq.(\ref{delta}), the numerical value decreases to 2.97 GeV$^{-1}$.
Furthermore, making use of the Rarita--Schwinger formalism and varying
the parameter $Z$, one can get sizeable changes in the $\Delta$
contributions ( e.g. $c_2^\Delta = 1.89, \, c_3^\Delta =-3.03 , 
\, c_4^\Delta = 1.42$ in GeV$^{-1}$ for $Z=-0.3$).
From this, we deduce the following ranges: $c_2^\Delta = 
1.9 \ldots3.8, \, c_3^\Delta =-3.8 \ldots -3.0 , 
\, c_4^\Delta = 1.4 \ldots 2.0$ (in GeV$^{-1}$).  

The Roper $N^* (1440)$ resonance
contributes only marginally,
\beqa  
c_2^R &=&  {g_A^2 m\tilde{R}\over 8 (m^{*2}-m^2)} =0.05 \, {\rm GeV}^{-1} \,\, ,
\nonumber \\
c_3^R &=& - {g_A^2 \tilde{R} \over 16 (m^*-m)} = -0.06 \, {\rm GeV}^{-1} \,\, ,
\nonumber \\    
c_4^R &=& {g_A^2 \tilde{R} \over 8 (m^*-m)} =0.12 {\rm GeV}^{-1} \, \, ,
\eeqa   
using $\tilde{R}=0.28$ as obtained from the partial decay width
$\Gamma (N^* \to N \pi ) \simeq 110\,$MeV \cite{bkmpipin}.

\medskip

Putting pieces together, we have for $c_2$, $c_3$ and $c_4$ from
resonance exchange (remember that $c_1$ was assumed to be saturated 
by scalar exchange)
\beqa \label{cireso}
c_2^{\rm Res} &=& c_2^\Delta + c_2^R =  3.83 + 0.05 = 3.88 \,\, , \nonumber \\
c_3^{\rm Res} &=& c_3^\Delta + c_3^S + c_3^R 
= -3.83 -1.40 - 0.06  = -5.29 \,\, , \nonumber \\
c_4^{\rm Res} &=& c_4^\Delta + c_4^\rho + c_4^R 
= 1.92 + 1.63 + 0.12 = 3.67 \,\, ,
\eeqa
with all numbers given in units of GeV$^{-1}$.
Comparison with the empirical values listed in table~1 shows that
these LECs can be understood  from resonance
saturation, assuming only that $c_1$ is entirely given by scalar meson
exchange. As argued before, the scalar meson parameters needed for
that are in good agreement with the ones derived from fitting NN
scattering data and deuteron properties  within the framework of a
one--boson--exchange model. We stress again that this $\sigma$--meson is an
effective degree of freedom which parametrizes the strong $\pi \pi$
 correlations (coupled to nucleons) in the scalar--isoscalar channel.
It should not be considered a novel degree of freedom which limits the
applicability of the effective field theory to a lower energy scale.
As pointed out before, there is some sizeable uncertainty related to the
$\Delta$ contribution as indicated by the ranges for the $c_i^{\rm Res}$
in table~1. It is, however, gratifying to observe that the empirical
values are covered by the band based on the resonance exchange model.

\medskip

The LECs $\krig \kappa_s=-0.12$ and $\krig \kappa_v = 5.83$ can be estimated 
from neutral vector meson exchange, in particular
\beq
\krig{\kappa}_s = \kappa_\omega \, , \quad \krig{\kappa}_v = \kappa_\rho
\,\,\, .
\eeq
Using e.g. the values from \cite{mmd}, $\kappa_\omega = -0.16 \pm
0.01$ and $\kappa_\rho = 6.1 \pm 0.4$, we see that the isoscalar and isovector
anomalous magnetic moments in the chiral limit can be well understood from 
$\omega$ and $\rho^0$ meson exchange. It is amusing that the isovector pion
cloud of the nucleon calculated to one loop allows to explain the observed 
difference between $\kappa_\rho$ and  $\kappa_v$. In strict vector meson
dominance these would be equal. It is well known \cite{hoeh} that the low 
energy part of the nucleon isovector spectral functions can not be 
understood in terms of the $\rho$--resonance alone.  


\section{Aspects of pion--nucleon scattering}

Having established that resonance saturation works rather well for the
dimension two LECs, we proceed to calculate the chiral expansion of
of the isovector  S--wave effective range parameter $b^-$ and of
the other two P--wave $\pi N$ scattering volumes up-to-and-including
terms of order $q^3$. Finally, we also work out the D-- and F--wave
threshold parameters $a_{l\pm}^\pm$, $l = 2,3$. Results for the subthreshold
parameters which do not receive any contribution from ${\cal L}_{\pi N}^{(3)}$
are collected in app.~B.

\subsection{Chiral expansion of P--wave scattering volumes}

We consider $P_2^+$, the isoscalar spin-flip scattering volume, 
and $P_1^-$ related to the isovector spin non-flip amplitude \cite{ew},
\begin{eqnarray} \label{P1val} 
P_1^- &=& \frac{4 \pi}{3}  ( 1 + \mu ) \Big(-2 a_{33}
-a_{31}+2a_{13}+a_{11} \Big) = (-2.52 \pm 0.03)\, M_{\pi}^{-3} \,, \\  
\label{P2val}
P_2^+ &=& \frac{4 \pi}{3} ( 1 + \mu) \Big(-2 a_{33} +2
a_{31}-a_{13}+a_{11} \Big) = (-2.74 \pm 0.03)\, M_{\pi}^{-3} \, .
\end{eqnarray}
Our aim is to see how well the empirical values given in
eqs.(\ref{P1val},\ref{P2val}) can be understood within chiral perturbation theory.
For that, we have to account for Born terms, one loop graphs and
insertions from ${\cal L}_{\pi N}^{(3)}$ (because of the crossing
properties of these amplitudes). In contrast to the previous cases,
the one--loop contributions are not finite and an appropriate
renormalization has to be performed.

Consider first the Born terms. Including all terms, in particular
the $\pi \pi \bar N N $ Weinberg vertex, the expansion to order $q^3$ gives
\begin{eqnarray}
 P_1^-({\rm Born}) &=& -{g_{\pi N}^2 \over 2 m^2}\biggl( {1\over
M_\pi} + {1\over m} + {3M_\pi \over 4 m^2} \biggr) +{1\over 4m F_\pi^2 }
\biggl( 1 + {M_\pi \over 2m} \biggr)  = -2.22 \, M_{\pi}^{-3} \\ 
 P_2^+({\rm Born}) &=& -{g_{\pi N}^2  \over 2 m^2}\biggl( {1\over M_\pi}
 +{1\over  m} +{M_\pi \over 4m^2} \biggr)= -2.29 \, M_{\pi}^{-3} \, ,
\end{eqnarray}
where the numbers refer to our standard set of parameters ($g_{\pi N}=13.4$). 
We now turn to the chiral loop corrections at order $q^3$. First, one has to
perform the standard coupling constant renormalization,
$\krig{g}_{\pi N} \to g_{\pi N}$. We use dimensional regularization
and the corresponding renormalization scale $\lambda$ is varied 
between $M_\rho = 0.77$ GeV and $m^*= 1.44$ GeV. In principle, this
scale--dependence would be balanced by the contribution from the LECs.
Since we use resonance saturation to estimate these, there remains a
small scale--dependent reminder which can not be fixed 
(compare also \cite{reso}). As a check on the one--loop calculation,
one verifies that the divergence appearing in $P_1^-$ is canceled by the
local counter term $O_1+O_2$ and 
the one in $P^+_2$ by the counter term $O_{15}$ of ref.\cite{eckerm}.
At the scale $\lambda = m$, we have 
\begin{eqnarray} 
 P_1^-({\rm Loop}) &=& 
-{M_\pi \over 48\pi^2 F_\pi^4}\biggl[\Big( 2 g_A^4 +5g_A^2+1\Big) \ln {M_\pi
\over \lambda}  + {1\over 3}g_A^4 +{7\over 2} g_A^2  + {1\over 2}\biggr]
\nonumber \\ &=& (0.25\pm 0.05) \, M_{\pi}^{-3} \,\,\, , \\   
P_2^+({\rm Loop}) &=& -{g_A^4
M_\pi   \over 24 \pi^2 F_\pi^4 }\biggl( {7\over 6} +\ln {M_\pi \over \lambda}
\biggr)= (0.05 \pm 0.02)\, M_{\pi}^{-3} \,\, . 
\end{eqnarray}
The uncertainty stems from the variation in $\lambda$ as described
above. The counter term contribution is estimated from
$\Delta$-resonance exchange employing the Rarita-Schwinger formalism,
\begin{eqnarray}  
P_1^-(\Delta) &=& {g_{\pi N}^2
M_\pi \over m^3 m_\Delta^2} \biggl[ {m_\Delta^2(m_\Delta-2m) \over 2(m_\Delta-
m)^2} +{2Z-1 \over 4} (m+m_\Delta)+Z^2m_\Delta \biggr] 
=  -0.35\, M_{\pi}^{-3} \,, \nonumber  \\ && \\  
P_2^+(\Delta) &=& { g_{\pi N}^2 M_\pi \over m^2 m_\Delta^2}
\biggl[-{m_\Delta m \over 2(m_\Delta-m)^2 } +{1\over 4} -Z^2 \biggr] = -0.33 \,
M_{\pi}^{-3} \,\, ,  
\end{eqnarray}
for the off--shell parameter $Z=-0.3$. Using the non-relativistic isobar model
and performing no chiral expansion one finds from the $\Delta(1232)$-resonance,

\begin{equation} P^-_1(\Delta) = P^+_2(\Delta) = - {g_{\pi N}^2 M_\pi \over
2m^2 [(m_\Delta-m)^2 - M_\pi^2 ] }= -0.58\, M_\pi^{-3}\,, \end{equation}
which is almost twice as large as before. Taking the average of both
$\Delta(1232)$-estimates and adding uncertainties in quadrature, the
chiral predictions to ${\cal O}(q^3)$ are
\beq \label{chp12}
P_1^- = (-2.44 \pm 0.13) \, M_{\pi}^{-3} \, \, , \quad
P_2^+ = (-2.70 \pm 0.12) \, M_{\pi}^{-3} \, \, .
\eeq
The major uncertainty comes here from the $\Delta(1232)$ contribution which
seems hard to pin down accurately. 
Further contributions at ${\cal O}(M_\pi)$ coming from the Roper 
resonance and the $\rho$-meson (as calculated
in \cite{Peccei}) fall into the error band given in eq.(\ref{chp12}).
We note that in both cases the Born terms
are dominant and $\Delta$ exchange amounts to a 18 and 17 \%
correction, respectively. The loop correction is very small for
$P_2^+$  and roughly $-10\%$ 
for $P_1^-$. Interestingly the small difference between $P_1^-$ and $P_2^+$
stems mainly from the chiral loops.  In tree level calculations
\cite{ew} and also the Skyrme soliton model \cite{HoSch}, 
$P_1^- - P_2^+ = 4\pi (1+\mu) (a_{13}-a_{31})$ is actually
zero as a consequence of $SU(4)$--spin--flavor symmetry.
 This quantity therefore serves as an interesting signal for chiral loop 
effects in the $\pi N$ P--wave amplitudes. Furthermore the chiral expansion of
these observables shows a good convergence (as expected for these particular
P--waves scattering volumes).  
The chiral predictions are well within the empirical values for $P_1^-$ and
$P_2^+$, however the theoretical uncertainty is larger than the experimental 
one. We conclude that also these particular $\pi N$ threshold
parameters can be understood within heavy baryon chiral perturbation theory. 

\subsection{The isovector S--wave effective range parameter}

The real part of the isovector forward scattering amplitude $T^-$
close to threshold takes a similar form than given in 
eqs.(\ref{expTp},\ref{betap}) for $T^+$, with
\beq
\beta^-  = \frac{1}{2 M_\pi} \frac{\partial}{\partial \omega}
{\rm Re} \, T^- (\omega) \bigl|_{\omega=M_\pi} =
4\pi(1+\mu) \, b^- + \frac{T^- (M_\pi)}{2 m M_\pi} + P_1^- \,\, .
\eeq
The second term is proportional to the isovector S--wave scattering length,
$a^-$, which we already discussed in detail in \cite{bkmpin2}. We just mention
that the prediction given in that paper agrees well with the recent
determinations from pionic atoms \cite{leisi}. Therefore, 
we will discuss here the isovector S--wave effective range parameter $b^-$,
which is smaller and of opposite sign than the isoscalar one, $b^- \simeq
-0.3\, b^+$. We prefer to keep the kinematical factor $4\pi(1+\mu)$ and thus
have to compare with the empirical value \cite{hoeh},
\beq 
4\pi (1+\mu) \, b^- = (0.19 \pm 0.09) \cdot M_\pi^{-3}\,.  
\eeq
The chiral expansion of this quantity takes the following form. From the tree
graphs one finds up to order ${\cal O}(M_\pi)$,
\beqa 
4\pi (1+\mu)\,b^-({\rm Born}) &=& {1\over 4 F_\pi^2 M_\pi} - {g_A^2 \over
2 m F_\pi^2} + {M_\pi \over 16 m^2 F_\pi^2} ( 2-5g_A^2) \nonumber \\
  &=& (0.57 - 0.30 - 0.02) \cdot M_\pi^{-3} = 0.25 \cdot M_\pi^{-3}\,,
\eeqa
with the contributions of the powers  $M_\pi^{-1,0,1}$ given separately. One 
sees that the truncation at order $M_\pi^0$ is already in agreement with the
experimental value (which has quite a large error bar). As a further
contribution we only mention the chiral loop correction. After renormalization
of the pion decay constant and the $\pi N$ coupling constant it reads,
\beq 
4\pi(1+\mu)\, b^-({\rm Loop}) = {M_\pi \over 48 \pi^2 F_\pi^4} \biggl[
(5g_A^2 -8) \ln{M_\pi \over \lambda} +{7\over 2} g_A^2 -1 \biggr] = 0.04 \cdot
M_\pi^{-3}\,, 
\eeq
for $\lambda =m$. Varying $\lambda$ between $M_\rho$ and $m^*$, this number
changes by less than 10 \%. The loop correction is smaller than the 
experimental uncertainty and this presumably holds for all other order 
$M_\pi$ counter term
contributions. We conclude, that also the value of the isovector S--wave 
effective range parameter $b^-$ can be understood within heavy baryon CHPT.  

\subsection{D-- and F--wave threshold parameters}

Finally, we discuss the eight D- and F-wave threshold parameters $a^\pm_{l
\pm}\,,\, l=2,3$. To these only the Born (lowest order tree) graphs and two
specific one loop graphs contribute at order $q^3$, but no counter terms from
${\cal L}_{\pi N}^{(2,3)}$. In the following expressions the loop 
contributions
are the ones carrying the factor $F_\pi^{-4}$. We find that in most cases the
chiral loop corrections are quite important to bring the 
chiral expansion close
to the experimental values. The latter are taken from \cite{hoeh}. 
We also remark that Moj\v zi\v s' calculation \cite{mm} 
of these threshold parameters is prior to ours. The
chiral expansion up to order $q^3$ reads:

\noindent D-wave threshold parameters: 
\begin{eqnarray} & &a^+_{2+} = - {g_A^2 (2+\mu) \over 120 \pi m F_\pi^2
M_\pi^2}+ {193 g_A^2 \over 115200 \pi^2 F_\pi^4 M_\pi} = -1.83 \cdot
10^{-3} M_\pi^{-5} \nonumber \\ & & a^+_{2+}({\rm exp}) = (-1.8 \pm 0.3) \cdot
   10^{-3} M_\pi^{-5}  \end{eqnarray}  
\begin{eqnarray} & &a^+_{2-} =  {g_A^2 (2+\mu) \over 480 \pi m F_\pi^2
M_\pi^2}+ {193 g_A^2 \over 115200 \pi^2 F_\pi^4 M_\pi} = 2.38 \cdot
10^{-3} M_\pi^{-5} \nonumber \\ & & a^+_{2-}({\rm exp}) = (2.2 \pm 0.3) \cdot
   10^{-3} M_\pi^{-5}  \end{eqnarray}  
\begin{eqnarray} & &a^-_{2+} = {g_A^2 (2+\mu) \over 120 \pi m F_\pi^2
M_\pi^2}+ {1+ g_A^2(7-5\pi) \over 14400 \pi^3 F_\pi^4 M_\pi} = 3.21 \cdot
10^{-3} M_\pi^{-5} \nonumber \\ & & a^-_{2+}({\rm exp}) = (3.2 \pm 0.1) \cdot
   10^{-3} M_\pi^{-5}  \end{eqnarray}  
\begin{eqnarray} & &a^-_{2-} = - {g_A^2 (2+\mu) \over 480 \pi m F_\pi^2
M_\pi^2}+ {2+ g_A^2 (14+15\pi)\over 28800 \pi^3 F_\pi^4 M_\pi} = -0.21 \cdot
10^{-3} M_\pi^{-5} \nonumber \\ & & a^-_{2-}({\rm exp}) = (0.1 \pm 0.2) \cdot
   10^{-3} M_\pi^{-5}  \end{eqnarray}  

\noindent F-wave threshold parameters:
\begin{eqnarray} & &a^+_{3+} = {g_A^2 \over 140 \pi  F_\pi^2 M_\pi^3} \biggl(
{1\over m^2} +{73  \over 5376 \pi F_\pi^2 } \biggr)  = 0.29 \cdot
10^{-3} M_\pi^{-7} \nonumber \\ & & a^+_{3+}({\rm exp}) = (0.42 \pm 0.13)
\cdot    10^{-3} M_\pi^{-7}  \end{eqnarray}  
\begin{eqnarray} & &a^+_{3-} = {g_A^2 \over 840 \pi  F_\pi^2 M_\pi^3} \biggl(
{73  \over 896 \pi F_\pi^2 } -{1\over m^2} \biggr)  = 0.06 \cdot
10^{-3} M_\pi^{-7} \nonumber \\ & & a^+_{3-}({\rm exp}) = (0.15 \pm 0.12)
\cdot    10^{-3} M_\pi^{-7}  \end{eqnarray}  
\begin{eqnarray} & &a^-_{3+} = {1 \over 140 \pi  F_\pi^2 M_\pi^3} \biggl(
{2+ g_A^2(18-7\pi) \over 3360 \pi^2 F_\pi^2 }-{g_A^2 \over m^2}  \biggr)  =
-0.20  \cdot 10^{-3} M_\pi^{-7} \nonumber \\ & & a^-_{3+}({\rm exp}) = (-0.25
\pm 0.02) \cdot    10^{-3} M_\pi^{-7}  \end{eqnarray}  
\begin{eqnarray} & &a^-_{3-} = {1 \over 840 \pi  F_\pi^2 M_\pi^3} \biggl(
{g_A^2\over m^2} +{3 + g_A^2 (27+14 \pi)  \over 840 \pi^2 F_\pi^2 } \biggr)  =
0.06  \cdot 10^{-3} M_\pi^{-7} \nonumber \\ & & a^-_{3-}({\rm exp}) = (0.10 \pm
0.02) \cdot    10^{-3} M_\pi^{-7}  \end{eqnarray}

\bigskip


\section*{Acknowledgements}

We are grateful to Martin Moj\v zi\v s for communicating his results prior to
publication.

\bigskip

\appendix
\section{Pion--nucleon scattering amplitude}
\def\theequation{\Alph{section}.\arabic{equation}}
\setcounter{equation}{0}

Here, we give explicit closed form expressions for  the one-loop 
contribution to 
the $\pi N$-scattering amplitude. In the center-of-mass (cms) frame the $\pi
N$-scattering amplitude $\pi^a(q) + N(p) \to \pi^b(q') + N(p')$ takes the
following form: 
\begin{equation} T^{ba}_{\pi N} = \delta^{ba} \Big[ g^+(\omega,t)+ i \vec
\sigma \cdot(\vec q\,'\times \vec q\,) \, h^+(\omega,t) \Big] +i \epsilon^{bac}
\tau^c \Big[ g^-(\omega,t)+ i \vec \sigma \cdot(\vec q\,'\times \vec q\,) \,
h^-(\omega,t) \Big] \end{equation}
with $\omega = v\cdot q = v\cdot q\,'$ the pion cms energy and $t=(q-q\,')^2$ 
the invariant momentum transfer squared. $g^\pm(\omega,t)$ refers to the
isoscalar/isovector non-spin-flip amplitude and $h^\pm(\omega,t)$ to the
isoscalar/isovector spin-flip amplitude. After renormalization of the pion
decay constant $F_\pi$ and the pion-nucleon coupling constant $g_{\pi N}$ one
finds the following one-loop contributions to the cms amplitudes
$g^\pm(\omega,t)$ and $h^\pm(\omega,t)$ at order $q^3$:
\begin{eqnarray} g^+(\omega,t)_{\rm loop} &=& {g_A^2 \over 32 \pi F_\pi^4}
\biggl\{ - {4\omega^2\over g_A^2} \sqrt{M_\pi^2-\omega^2} +
(M_\pi^2-2t) \biggl[
M_\pi +  {2M_\pi^2-t \over 2 \sqrt{-t}} \arctan{\sqrt{-t}\over2M_\pi} \biggr]
\nonumber \\ & & \qquad \qquad + {4g_A^2 \over 3 \omega^2}(2\omega^2 +t
-2M_\pi^2) \Big[(M_\pi^2-\omega^2)^{3/2}-M_\pi^3 \Big] \biggr\} \end{eqnarray}

\begin{eqnarray} h^+(\omega,t)_{\rm loop} &=& {g_A^4 \over 24\pi^2 F_\pi^4}
\biggl\{ - \omega \biggl( \ln{M_\pi \over \lambda} +{1\over6} \biggr) -
{M_\pi^2 \over \omega } + {(M_\pi^2-\omega^2)^{3/2} \over \omega^2}
\arcsin{\omega \over M_\pi} \biggr\} \end{eqnarray}
\begin{eqnarray} g^-(\omega,t)_{\rm loop} &=& {\omega\over 48 \pi^2 F_\pi^4}
\biggl\{ 3 \omega^2 \biggl( 1 -2 \ln {M_\pi \over \lambda} \biggr)-6 \omega
\sqrt{M_\pi^2 - \omega^2} \arcsin{\omega \over M_\pi} \nonumber \\ & & + g_A^4
(2 \omega^2 +t -2M_\pi^2) \biggl[ {5\over6} - \ln{M_\pi \over \lambda}
-{M_\pi^2 \over \omega^2} +{(M_\pi^2-\omega^2)^{3/2} \over \omega^3}
\arcsin{\omega \over M_\pi} \biggr] \nonumber \\ & & + \biggl[ 2 M_\pi^2
(1+2g_A^2) - {t\over 2} (1+5g_A^2) \biggr] \sqrt{1- {4M_\pi^2 \over t}}\, \ln
\biggl({\sqrt{4M_\pi^2-t}+\sqrt{-t}\over 2M_\pi}\,\biggr) \nonumber \\ & & 
-{t\over 2} (1+5g_A^2) \ln{M_\pi \over \lambda} +{t\over 12} (5 + 13 g_A^2) -
2M_\pi^2 (1+2g_A^2) \biggr\} \end{eqnarray}

\begin{eqnarray} h^-(\omega,t)_{\rm loop} &=& {g_A^2 \over 32\pi F_\pi^4}
\biggl\{ - M_\pi +{t-4M_\pi^2 \over 2 \sqrt{-t}} \arctan{\sqrt{-t} \over
2M_\pi}  + {4 g_A^2 \over 3 \omega^2} \Big[ (M_\pi^2-\omega^2)^{3/2}-M_\pi^3
\Big]  \biggr\} \nonumber \\ && \end{eqnarray} 
The analytic continuation above threshold $\omega > M_\pi$ is done via the
formulae 
\begin{equation} \sqrt{1-x^2} = -i \sqrt{x^2-1}\, , \qquad \arcsin x = {\pi
\over 2} + i \, \ln(x+\sqrt{x^2-1})\quad .  \end{equation}

The $t$-dependences of the loop-amplitudes $g^\pm(\omega,t)_{\rm loop}$ and
$h^\pm(\omega,t)_{\rm loop}$ show an interesting structure, if one discards
terms proportional to $g_A^4$. The $t$-dependence of
$h^+(\omega,t)_{\rm loop}$ is then given by $(2t-M_\pi^2)/(3M_\pi^2F_\pi^2)\,
\sigma(t)_{\rm loop}$, with $\sigma(t)$ the nucleon scalar form
factor. Furthermore, the $t$-dependence of $g^-(\omega,t)_{\rm loop}$ becomes
equal to $\omega/(2F_\pi^2)\,G_E^V(t)_{\rm loop}$, with $G_E^V(t)$ the nucleon
isovector electric form factor (normalized to unity). Finally, $h^-(\omega,t)_{
\rm loop}$ has the same $t$-dependence as $-1/(4mF_\pi^2)\,  G_M^V(t)_{\rm
loop}$, with $G_M^V(t)$ the  nucleon isovector magnetic form factor. The
one-loop calculation of these nucleon form factors can be found in [2].

\section{Results for some subthreshold parameters}
\def\theequation{\Alph{section}.\arabic{equation}}
\setcounter{equation}{0}

Here, we collect the results for those coefficients of the subthreshold
expansion (around $\nu = t = 0$) which to order $q^3$ are pure loop
effects. The experimental values are taken from \cite{hoeh}.

\renewcommand{\arraystretch}{1.3}

\begin{center}

\begin{tabular}{|c|c|c|}
    \hline
    Quantity      & One loop result   &  Experimental value \\
    \hline
$d^+_{11}$ &  $g_A^4 /(64 \pi F_\pi^4 M_\pi) = 0.08 \,
M_\pi^{-5}$  & $(0.17 \pm 0.01)\, M_\pi^{-5} $ \\

$d^+_{20}$ &  $ (12+5g_A^4)/( 192 \pi F_\pi^4 M_\pi) = 0.235 \,
M_\pi^{-5}$ & $(0.200 \pm 0.005)\, M_\pi^{-5}$ \\

$d^+_{02}$ & $ 193g_A^2 /( 15360 \pi F_\pi^4 M_\pi) = 0.036 \,
M_\pi^{-5}$ & $(0.036 \pm 0.003)\, M_\pi^{-5}$ \\

$b^+_{01}$ & $  0 $ & $ (0.18 \pm 0.01)\, M_\pi^{-5}$ \\

$b^+_{10}$ & $  g_A^4 m /( 60 \pi^2 F_\pi^4 M_\pi^2) = 0.18 \,
M_\pi^{-5}$ & $  (-1.00 \pm 0.02)\, M_\pi^{-5}$ \\

$b^+_{11}$ & $   0 $ & $ (0.08 \pm 0.01)\, M_\pi^{-7}$ \\

$b^+_{20}$ & $  g_A^4 m /( 210 \pi^2 F_\pi^4 M_\pi^4) = 0.05 \,
M_\pi^{-7}$ & $ (-0.31 \pm 0.02)\, M_\pi^{-7}$ \\

$b^+_{02}$ & $   0$ & $  -0.01\, M_\pi^{-7}$ \\

$d^-_{11}$ & $ g_A^4/( 240 \pi^2 F_\pi^4 M_\pi^2) = 0.007 \,
M_\pi^{-6}$ & $  (-0.042 \pm 0.003)\, M_\pi^{-6}$ \\

$d^-_{20}$ & $  (7+g_A^4)/(168 \pi^2 F_\pi^4 M_\pi^2 )=0.032 \,
M_\pi^{-6}$ & $ (-0.039 \pm 0.002)\, M_\pi^{-6}$ \\

$d^-_{02}$ & $  (1+7g_A^2) /(1920\pi^2 F_\pi^4 M_\pi^2 )=0.004 
\,M_\pi^{-6}$ & $ (0.010 \pm 0.001)\, M_\pi^{-6}$ \\

$b^-_{01}$ & $g_A^2m/(96 \pi F_\pi^4 M_\pi )=0.20 \,
M_\pi^{-4}$ & $ (0.24 \pm 0.01)\, M_\pi^{-4}$ \\

$b^-_{10}$ & $  g_A^4m /(32 \pi F_\pi^4 M_\pi )=1.06 \,
M_\pi^{-4} $ & $(1.08 \pm 0.05)\, M_\pi^{-4}$ \\

$b^-_{11}$ & $ 0$ & $(-0.055 \pm 0.005)\, M_\pi^{-6}$ \\

$b^-_{20}$ & $ g_A^4m/( 192 \pi F_\pi^4 M_\pi^3)=0.18 \,
M_\pi^{-6} $ & $(0.29 \pm 0.02)\, M_\pi^{-6}$ \\

$b^-_{02}$ & $   g_A^2m/(1920 \pi F_\pi^4 M_\pi^3 )=0.010 \,
M_\pi^{-6}$ & $ (0.025 \pm 0.002)\, M_\pi^{-6}$ \\
   \hline
  \end{tabular}
\smallskip
\end{center}
\bigskip

Obviously, only in some cases the one-loop result is in good agreement with
the empirical values as deduced from the Karlsruhe--Helsinki (KH) phase shift 
analysis. Note, however, that recent low energy $\pi N$-scattering data 
from PSI \cite{joram}
show some disagreement with the KH80 solution of $\pi N$ dispersion 
analysis. It therefore seems necessary to redo the $\pi N$-dispersion analysis
with the inclusion of these new data. A new determination of the subthreshold
coefficients is now also called for. 



\end{document}